# Controllable highly oriented skyrmion track array in Fe$_3$GaTe$_2$

Yunhao Wang[1,2,*], Shiyu Zhu[1,2,*,†], Chensong Hua[3,*], Guojing Hu[1], Linxuan Li[2], Senhao Lv[1], Jianfeng Guo[1], Jiawei Hu[1,2], Runnong Zhou[1,2], Zizhao Gong[1], Chengmin Shen[1,2], Zhihai Cheng[4], Jinan Shi[2], Wu Zhou[2], Haitao Yang[1,2], Weichao Yu[3,‡], Jiang Xiao[3], Hong-Jun Gao[1,2,5,§]

[1] Beijing National Center for Condensed Matter Physics and Institute of Physics, Chinese Academy of Sciences, 100190 Beijing, China.

[2] School of Physical Sciences, University of Chinese Academy of Sciences, 100190 Beijing, China.

[3] Department of Physics and State Key Laboratory of Surface Physics, Fudan University, Shanghai 200433, China

[4] Department of Physics and Beijing Key Laboratory of Optoelectronic Functional Materials & Micro- nano Devices, Renmin University of China, Beijing 100872, China

[5] Hefei National Laboratory, 230088 Hefei, Anhui, China.

*These authors contributed equally to this work.
†syzhu@iphy.ac.cn (S. Z.)
‡wcyu@fudan.edu.cn (W. Y.)
§hjgao@iphy.ac.cn (H.-J. G.).

# ABSTRACT

Magnetic skyrmions are emerging as promising candidates for next-generation information technologies, while the realization of scalable skyrmion lattices with tailored configurations is essential for advancing fundamental skyrmion physics and developing future applications. Here we achieved the controllable generation and regulation of a large-area, highly oriented skyrmion track array (STA) in ferromagnetic $Fe_3GaTe_2$ using a vector magnetic field manipulation technique. The orientation and ordering of STA, along with the types and density of skyrmions, are precisely controlled by modulating parameters during the manipulation. The critical roles of in-plane magnetic fields and Dzyaloshinskii–Moriya interaction in STA generation is further confirmed by micromagnetic simulation. Our findings develop a strategy for engineering large-area and highly-oriented skyrmion configurations, offering a new pathway for the future application of next-generation spintronic and information technologies.

# I. INTRODUCTION

Magnetic skyrmions, the topologically protected swirling spin structures, are considered promising candidates for energy-efficient nanoscale memory and logic devices [1]. The observation of skyrmions has been reported across various platforms including bulk chiral magnet crystals [2–5], exfoliated thin flakes [6], interfacially asymmetric magnetic multilayers [7–10], and nanowires [11,12]. Recently, magnetic van der Waals (vdW) crystals have emerged as a new arena for exploring novel magnetic properties and hold significant potential for ultra-compact spintronic devices, owing to their easy exfoliation process and capability to maintain long-range ferromagnetic order down to atomic layers or thin flakes [6,13,14]. Specifically, strong perpendicular magnetic anisotropy (PMA), dipolar interaction, Dzyaloshinskii–Moriya interaction (DMI) and skyrmion lattices have been reported in an above-room-temperature vdW ferromagnet $Fe_3GaTe_2$ [15–23], providing a new platform for manipulating and constructing skyrmion-based configurations.

Currently, skyrmions are commonly reported to be generated from labyrinthine or stripe domains under a perpendicular magnetic field [10,24,25], while the generation on certain materials has also been reported using various other stimuli such as current [7], laser [26], X-ray [27], electron beam [28,29] and spin-polarized tunneling current [30]. These techniques typically result in either a pure skyrmion phase or a hybrid phase comprising skyrmions interspersed with disordered labyrinthine domains. Recently advancements have demonstrated current-driven skyrmion motion in devices [31–33], stripe domains [34], and domain walls [35], underscoring the significance of configurations that confine skyrmion chains into one-dimensional structures. Accordingly, developing innovative strategies to design and fabricate ordered magnetic structures that integrate skyrmion chains into well-defined racetracks has emerged as a pivotal research objective for both fundamental studies and potential applications.

Here we report the generation and regulation of a large-area, highly oriented skyrmion track array (STA) in ferromagnet $Fe_3GaTe_2$, achieved through a vector magnetic field modulation strategy. The highly oriented STA are observed extending across hundreds of micrometers by using magnetic force microscopy (MFM). It has been confirmed that the orientation, ordering, and skyrmion density of the STA are rigorously controlled by the vector magnetic field in our strategy. Additionally, two types of skyrmions with distinct characteristics and magnetic evolution paths have been controllably generated within the STA. We demonstrate that the underlying mechanism for the formation of STA is highly dependent on the shrinking propagation period of stripe domains induced by a varying in-plane magnetic field, as supported by micromagnetic simulation.

## II. HIGHLY ORIENTED SKYRMION TRACK ARRAY ON $Fe_3GaTe_2$

$Fe_3GaTe_2$ is a vdW ferromagnet with each layer consisting of a Fe/FeGa/Fe heterometallic slab enclosed between two Te layers [Fig. 1(a)]. Single crystals of $Fe_3GaTe_2$ are grown using chemical vapor transport (CVT) method and their high quality is validated by various characterization techniques (Figs. S1(a)-S1(f)) including X-ray diffraction (XRD), energy-dispersive X-ray spectroscopy (EDS), and aberration-corrected scanning transmission electron microscopy (STEM). Magnetization measurements reveal a high Curie temperature ($T_c$) at ~356 K, and a saturation field of around 0.6 T along the $c$-axis (Figs. S1(g)-S1(i)). The strong PMA of $Fe_3GaTe_2$ crystal is verified by the unsaturated $M$-$H$ curve in the $ab$-plane high magnetic field up to 5 T.

The initial magnetism of $Fe_3GaTe_2$, cooled from room temperature to 4 K without an external magnetic field, is characterized by MFM measurements. These measurements reveal smooth but disordered labyrinthine domains interspersed with a few skyrmions distributed dispersedly among them (Fig. S2). As the out-of-plane magnetic field ($B_\perp$, $B_z$) increases, the labyrinthine domains and skyrmions gradually shrink and ultimately vanish at $B_\perp \approx 0.6$ T, which aligns well with the saturation field observed in the $M$-$H$ measurement results. By contrast, fewer skyrmions are observed among the labyrinthine domains when magnetic field is decreased back to zero.

To enhance the ordering and density of skyrmions, we have successfully engineered a highly oriented STA using a vector magnetic field strategy (see Figure 1(b) and movie S1 in Supplemental Materials). This macro manipulation technique enables uniform and precise control over the orientation, ordering, and density of the STA across the entire sample, extending over several millimeters. The high-density skyrmion lattice, regulated by the highly oriented stripe domains, is separated into aligned skyrmion chains on the $Fe_3GaTe_2$ surface, as demonstrated in an area exceeding 166 μm × 147 μm [Fig. 1(c)].

In this vector magnetic field strategy, the magnetic configuration is initially reset using an out-of-plane magnetic field $B_\perp$ exceeding the saturation field (such as $B_\perp = 1.0$ T in our demonstration) [Fig. 1(d), step ①]. Subsequently, $B_\perp$ is reduced to 0.6 T (step ②), which remains well above the thresholds for labyrinthine domain emergence in Sample #1 ($B_\perp = 0.20$ T, as shown in Figure S2) and Sample #2 ($B_\perp = 0.35$ T, as shown in Figure S3). Under the strong magnetic field $B_\perp$ ( step ① and step ②), $Fe_3GaTe_2$ bulk maintains a saturated magnetization state, with the magnetization $m$ fully points along the field direction. As $B_\perp$ is gradually decreased ( step ③ to step ⑤), the magnetic moment in some regions begins to reverse, forming magnetic domains as the total magnetization decreases, which is confirmed by the evolution of normalized magnetization throughout the process (Figs S4(c)). Meanwhile, the applied in-plane magnetic field $B_\parallel$ facilitates the transformation of these emerging domains into well-aligned, highly oriented stripes [Figs. 1(e) and 1(f)]. Increasing the in-plane magnetic field to $B_\parallel^* = 3.0$ T then generates fragmented structures between adjacent stripes [Fig. 1(g), step ⑥], which remain stable when the field is decreased to zero [Fig. 1(h), step ⑦]. The mechanism underlying this novel phenomenon will be discussed in details in the following text. These fragmented structures between the stripes further evolve into smooth skyrmion chains, arranged alternately with stripes, upon the reapplication of an out-of-plane field [Fig. 1I, step ⑧]. Further detailed information on the magnetic domain evolution can be found in Supplemental Material (Figs S4 and S5).

We also examined the magnetic stability of the highly oriented STA by reducing the magnetic field to zero and subsequently transferring $Fe_3GaTe_2$ crystal into an atmospheric environment at room temperature. The precisely controlled characteristics of the highly oriented STA remained stable, as demonstrated over an area exceeding 80 μm × 100 μm (Fig. S6), highlighting its potential applicability in skyrmion-based spintronic devices.

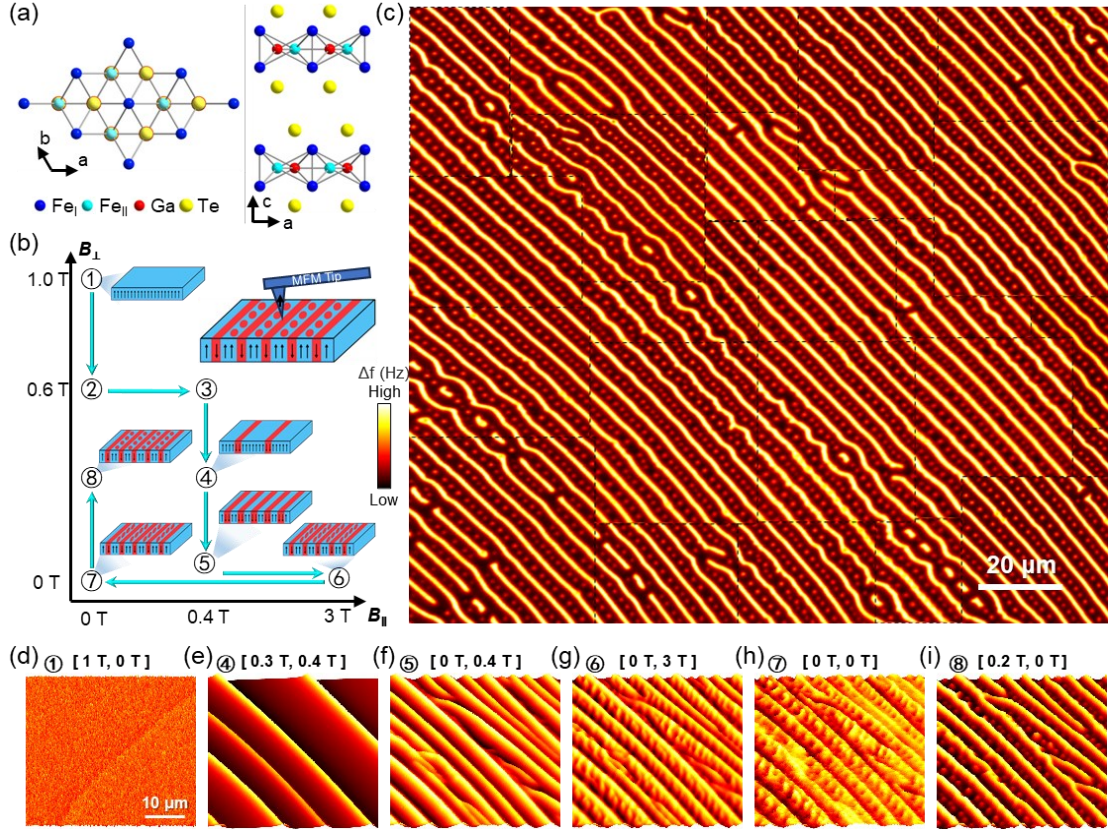

FIG. 1. Highly oriented skyrmion track array (STA) on $Fe_3GaTe_2$. (a) Crystal structure of $Fe_3GaTe_2$. (b) Schematic illustrating the generation of a highly oriented STA using a vector magnetic field strategy. Inset: Schematic of MFM measurements. (c) A stitched large area MFM image of ordered STA (166 μm × 147 μm) in step ⑧ ($B_\perp$ = 0.3 T, $B_\parallel$ = 0 T). In-plane magnetic field $B_\parallel'$ hosts an in-plane direction of $\theta_{xy}$ = 45° in this case. (d)-(i) MFM images unveiling magnetic domain evolution in the formation of STA, which correspond to different steps of magnetic field manipulation procedure in (b). The specific step and magnetic field conditions [$B_\perp$, $B_\parallel$] for each image are indicated.

## III. REGULATION OF STA VIA VECTOR MAGNETIC FIELD

The orientation of the STA is easily regulated by adjusting the directions of in-plane magnetic field. We performed a series of vector magnetic field manipulation during the formation process of stripe domains (steps ③-⑤), varying the direction ($\theta_{xy}$) of the in-plane magnetic field $B_\parallel'$ with its strength fixed at 0.4 T. In step ③, the applied $B_\parallel'$ induces an integral canting of magnetic moment toward its direction. Since the out-of-plane component of magnetization remains saturated and spatially uniform, no signal contrast appears in MFM images. When $B_\perp$ decreases below the saturation field, uniformed magnetization breaks down as total magnetization decreases, resulting in the appearance of magnetic domains in the MFM images. To regulate the domain direction, MFM measurements were captured with $B_\perp$ set to 0.2 T and $B_\parallel'$ held constant at 0.4 T, while the in-plane direction was altered [Fig. 2(a)]. The stripe domains exhibit a highly adjustable and precisely controllable orientation, consistently aligning perpendicular to $B_\parallel'$, indicating the presence of Bloch-type magnetic structures. The Bloch-type domain structure is supported by the orientation of simulated stripe domains, which shows a perpendicular direction with $B_\parallel$ for Bloch type and a parallel direction with $B_\parallel$ for Néel type (Fig. S7). The propagation direction of the stripes ($\theta_s$) and the ordering of the stripe domains were further analyzed using fast Fourier transform (FFT) results [Fig. 2(b)], which demonstrate excellent alignment with the in-plane field direction $\theta_{xy}$.

This orientation regulation also applies to skyrmion chains. A series of measurements of final STA structure (step ⑧) reveal that the skyrmion chains are always aligned and sandwiched between the pre-constructed stripe domains [Fig. 2(c)]. The FFT results confirm the alignment of the skyrmion chains, showing a more regular FFT pattern with intensity concentrated around two points, which indicates more uniform spacing between the stripes compared to pure stripe phase (step ④) [Fig. 2(d)].

The alignment and ordering of the stripes and skyrmion chains are quantified through directional statistics derived from FFT results. Both the integrated intensity ($I_{FFT}$, red dots) and the shape parameter ($S_{FFT}$, blue dots) are plotted as functions of the in-plane direction $\theta$ [Figs. 2(e)-2(f)]. The integrated intensity $I_{FFT}$ is calculated by integrating the intensity from the $\Gamma$ point to a suitable length that encompasses the main features of the FFT image at a specific in-plane direction $\theta$. The shape parameter $S_{FFT}$ is determined as the distance from the $\Gamma$ point to the edge at a given $\theta$, using a two-dimensional differentiation method. Both the integrated intensity curves and shape parameter curves exhibit clear and consistent peak characteristics with a periodicity of $\pi$ at both aligned stripe phase (step ④) and STA phase (step ⑧), providing a quantified view of the stripe propagation direction $\theta_s$ in each procedures. The strong linear relationship between the in-plane field direction $\theta_{xy}$ and the stripe propagation direction $\theta_s$ reveals the ability to precisely and continuously manipulate the STA direction [Fig. 2(g)]. Detailed information on orientation regulation is available in the complete MFM images and corresponding FFT images (Figs S8 and S9).

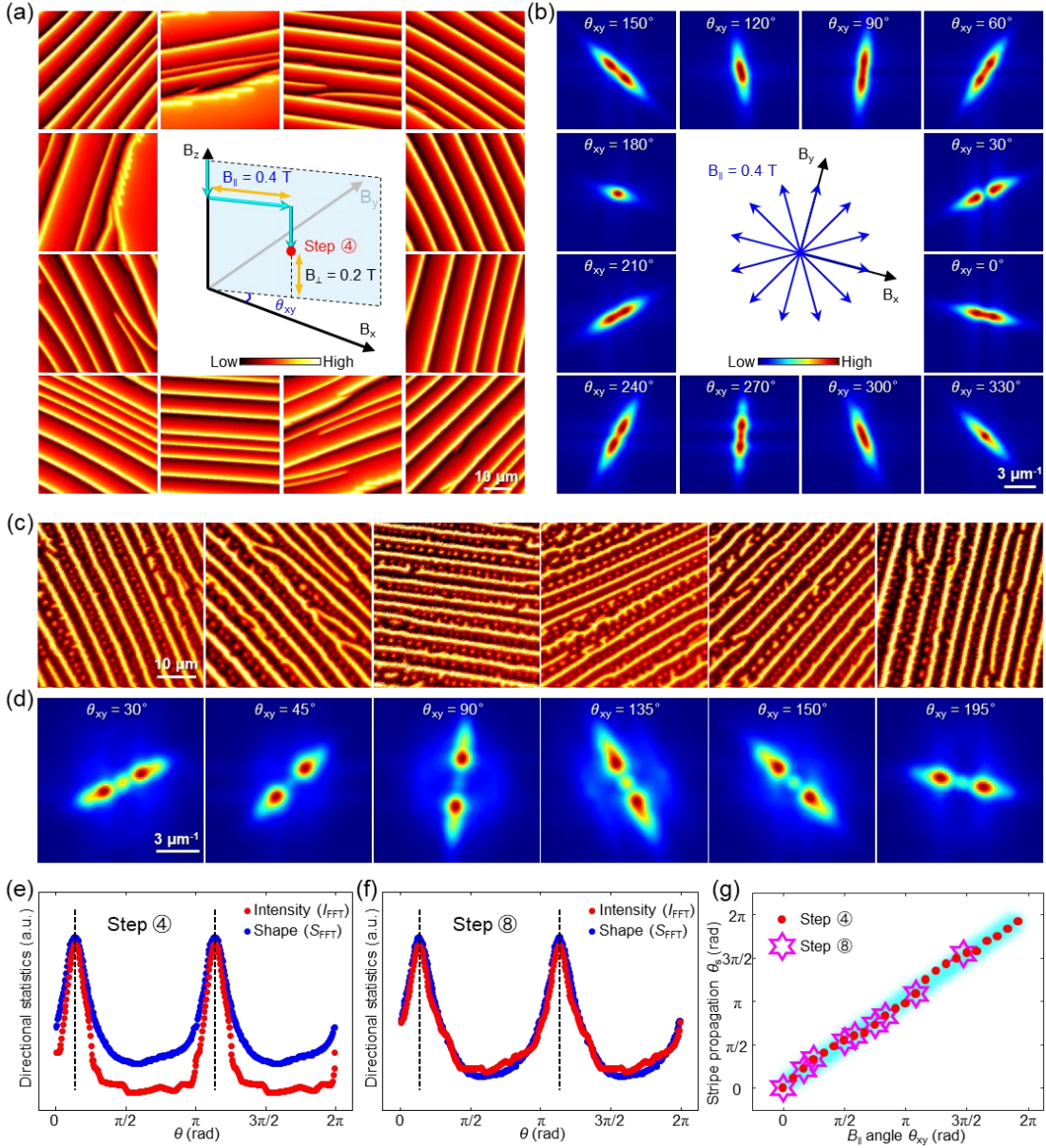

FIG. 2. Regulation of STA orientation by vector magnetic field. (a) Stripe domains in step ④ ($B_\perp$ = 0.2 T, $B_\parallel$ = 0.4 T) regulated by varying direction of in-plane magnetic field. Inset: A schematic illustrating the vector magnetic field manipulation procedure, where the angle $\theta_{xy}$ varies across different procedures, steps by $\pi/6$ from 0 to $2\pi$. (b) FFT results of MFM images in (a). (c) Highly oriented STA in step ⑧ ($B_\perp$ = 0.2 T, $B_\parallel$ = 0 T) regulated by varying direction of in-plane magnetic field, with the $\theta_{xy}$ = 30°, 45°, 90°, 135°, 150°, and 195°. (d) FFT results of MFM images in (c). (e),(f) An example summarizing the variation of $I_{FFT}$ and $S_{FFT}$ with $\theta$ in the FFT results, at $\theta_{xy}$ = 30° during step ④ and step ⑧. Both data display consistent peak characteristics, with the peak position indicating the propagation direction of the stripes ($\theta_s$). (g) Variation of the extracted peak position as a function of the in-plane magnetic field direction $\theta_{xy}$. The values extracted from the $I_{FFT}$ are shown as representatives, with the $\theta_s$ values for stripes in step ④ marked by red dots and those for the STA in step ⑧ marked by purple hexagonal stars. The shadow area indicates the trend of the linear relationship.

The ordering of stripe domains is visibly dependent on, and can be effectively controlled by, the value of the in-plane field $B_\parallel'$. We demonstrate this dependance by varying $B_\parallel'$ from 0 T (red) to 0.2 T (purple), 0.45 T (blue), and 1.0 T (green), applying these values in step ③ and assessing their effects in stripe phase (step ④) and STA phase (step ⑧) [Fig. 3(a)]. In independent vector magnetic field manipulating process with varying $B_\parallel'$, the stripe domains (step ④) exhibit distinct ordering as $B_\parallel'$ increases, with stripes aligning in the same direction, bending and branching structures disappearing, and more uniform stripe spacing developing across the different values of $B_\parallel'$ [Fig. 3(b)]. The skyrmion chains in STA structure (step ⑧) display similar features to the stripes in step ④, reflecting the strong influence of the in-plane magnetic field $B_\parallel'$ applied in step ③ [Fig. 3(c)]. Complete procedures of these magnetic field operations are detailed in the MFM images (Figure S10).

The impact of $B_\parallel'$ on stripe ordering and STA is quantified by statistical analysis of the magnetic structure. The order factors of stripe domains in step ④ ($B_\perp$ = 0.2 T) are estimated using the aspect ratios $R_I$ and $R_S$, derived from quotient of maximum and minimum values of integrated intensity curves (extracted from $I_{FFT}$, red dots) and shape parameter curves (extracted from $S_{FFT}$, blue dots). Both methods reveal a strong positive correlation between stripe ordering and varying $B_\parallel'$ [Fig. 3(d)]). The order factors rise gradually with increasing $B_\parallel'$, but show slower growth and increased fluctuations when $B_\parallel'$ exceeds 0.4 T, corresponding to the disappearance of bending and branching structure [Fig. 3(b)]. As expected, the final density of stripe domains at $B_\perp$ = 0 T (step ⑤) increases with $B_\parallel'$ and is saturated when $B_\parallel'$ exceeds 0.4 T [Fig. 3(e), blue squares]. However, the density of skyrmions exhibits a non-monotonic relationship, unexpected parabolic trend, peaking between 0.4 and 0.6 T [Fig. 3(e), red hexagonal stars]. For $B_\parallel'$ < 0.4 T, curled stripes squeeze the space available for skyrmions, while for $B_\parallel'$ > 0.4 T, fine branches rather than skyrmions are observed, both conditions leading to decreased skyrmion density.

In addition to $B_\parallel'$, the parameter $B_\parallel^*$ applied to induce precursors of skyrmions (step ⑥) is critical for controlling skyrmion density. We performed a series of independent vector magnetic field operations with various $B_\parallel^*$, specifically $B_\parallel^*$ = 0.4 T, 1 T, 2 T, and 3 T [Fig. 3(f)]. MFM images of ultimate STA structures (step ⑧) reveal an increase in skyrmion density with different $B_\parallel^*$ values [Fig. 3(g)]. To further understand the influence of $B_\parallel^*$ after the formation of aligned stripes, we studied the magnetic structure during $B_\parallel^*$ application (from step ⑤ to step ⑥) [Fig. 3(h)]. Gradually increasing $B_\parallel^*$ from 0.4 T to 3 T reveals many small branching structures emerging from the stripe domains, which then become spatially separated and lay the groundwork for the skyrmions in STA by providing the initial magnetic topology [36].

The regulation of skyrmion density by $B_\parallel^*$ is further quantified by statistical analysis under various $B_\parallel^*$ values on two samples, demonstrating a strong positive relationship on both samples [Fig. 3(i)]. It is important to note that, the stipe density is not affected by $B_\parallel^*$, but rather sets an upper limit on the number of skyrmion chains sandwiched between the stripes. Consequently, the increasing skyrmion density is primarily due to the reduced spacing among skyrmions within each chain. As expected, histograms summarizing skyrmion spacing reveal a negative relationship with $B_\parallel^*$ [Fig. 3(j)], with spacing at $B_\parallel^*$ = 0.4 T being notably larger compared to other values.

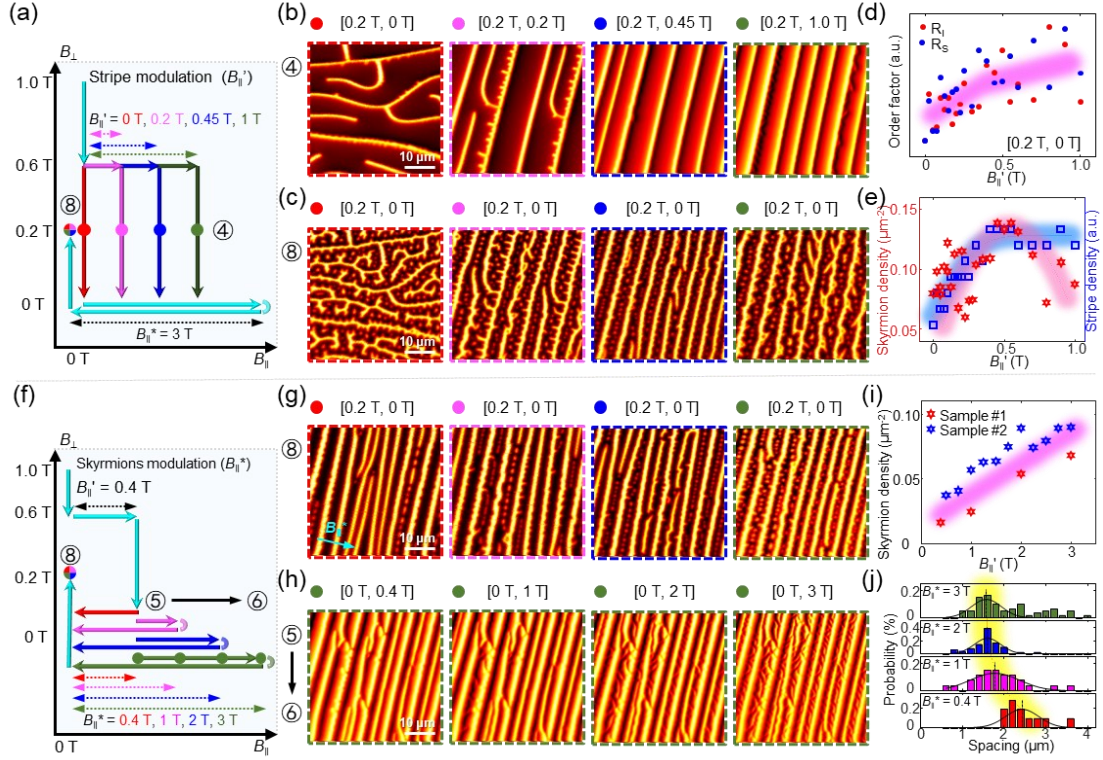

FIG. 3. Modulation of STA ordering and skyrmion density. (a) Schematics of ordering modulation for stripe domains in step ④ and STA in step ⑧ with varying $B_\parallel'$. (b) Labyrinthine/stripe domains induced in step ④ by variable $B_\parallel'$. (c) Sequent STA induced in step ⑧. (d) Variation of stripe order factors extracted in step ⑧ ($B_\perp = 0.2$ T, $B_\parallel = 0$ T), tuned by increasing $B_\parallel'$, $\theta_{xy} = 0°$. (e) Variation of stripe density (blue squares) in step ⑤ ($B_\perp = 0.2$ T, $B_\parallel = B_\parallel'$) and skyrmion density (red hexagonal stars) in step ⑧ ($B_\perp = 0.2$ T, $B_\parallel = 0$ T), under increasing $B_\parallel'$, $\theta_{xy} = 0°$. (f) Schematics of modulation of skyrmion density with $B_\parallel^*$. (g) STA with rising skyrmion density under increasing value of $B_\parallel^*$. (h) Magnetic domain evolution from step ⑤ to step ⑥, with $B_\parallel$ increasing from 0.4 T to 3 T. (i) Data from two samples showing skyrmion densities in step ⑧ ($B_\parallel = 0$ T, $B_\perp = 0.2/0.3$ T for Sample #1/#2, respectively) manipulated by $B_\parallel^*$. (j) Histograms summarizing skyrmion spacing distributions in step ⑧ ($B_\perp = 0.2$ T, $B_\parallel = 0$ T) with various $B_\parallel^*$. The color-coded manipulation, specific steps, and magnetic field conditions [$B_\perp$, $B_\parallel$] for each image are indicated in (b), (c), (g) and (h). The shadow areas indicate the trend of the relationships in (d), (e), (i) and (j).

# IV. TWO TYPES OF SKYRMIONS AND MECHANISMS OF STA GENERATION

Focusing on the behavior of skyrmions in STA (step ⑧), we compare the generation process of skyrmions under different in-plane magnetic fields, specifically $B_\parallel^* = 0.4$ T, 1 T, and 3 T, using MFM measurements in stripe phase (step ⑤) and STA phase (step ⑧) [Fig. 4(a)]. The independent vector magnetic field manipulation procedures reveal the evolution paths of two distinct types of skyrmions [Fig. 4(b)]: Skyrmions type-I (Sk-I) generated at lower $B_\parallel^*$ values and skyrmions type-II (Sk-II) generated at higher $B_\parallel^*$ values. Procedure with $B_\parallel^* = 0.4$ T generates Sk-I, where skyrmions arise from the breakdown of stripe domains (cyan boxes). The transformation from stripe domains to skyrmions typically occurs in thin stripes or weaker parts of stripe domains, which are more fragile under increasing $B_\perp$, and are transformed into skyrmion chains preferentially. Procedure with $B_\parallel^* = 3$ T generate Sk-II (purple boxes), where skyrmions originate from newly formed small branching magnetic structures sandwiched between stripe domains as previously discussed [Fig. 3(h)], resulting in higher skyrmion density. Procedure with $B_\parallel^* = 1$ T generate both types of skyrmions. The observed skyrmion types offer a consistent understanding with the sudden change skyrmion spacing with varying $B_\parallel^*$ [Fig. 3(j)].

The 3D image [Fig. 4(c)] and line profiles [Fig. 4(d)] provide a comparative overview of the two skyrmion types, highlighting differences in Δf signal strength, skyrmion size, and spacing within skyrmion chains. The distribution of MFM signal $\Delta f_{sk}$ (Δf contrast around a single skyrmion) of two types of skyrmions shows that, Sk-I exhibits a significantly higher $\Delta f_{sk}$ signal (97.7 Hz) compared to Sk-II (60.5 Hz), indicating stronger out-of-plane magnetization [Fig. 4(e)]. Additionally, extracted full width at half maximum (FWHM) [Fig. 4(f)] reveals a larger size of Sk-I (1.35 μm) compared to Sk-II (0.84 μm). The separation within skyrmion chains, shows a spacing of around 2.49 μm for Sk-I, larger than 1.70 μm for Sk-II [Fig. 4(g)], consistent with the higher skyrmion density for Sk-II and the decreasing spacing for larger $B_\parallel^*$. These differences suggest distinct three-dimensional magnetic structures for Sk-I and Sk-II.

We further investigated the evolution of Sk-I and Sk-II with increasing out-of-plane magnetic field $B_\perp$ during step ⑧. Increasing $B_\perp$ from 0.05 T to 0.4 T causes the $\Delta f_{sk}$ of

Sk-I to drop to about one-third of its original value, while Sk-II remains relatively constant, leading to close values of both types before vanishing from the measurement at $B_\perp$ = 0.45 T [see Fig. 4(h) and detailed images in Fig. S11]. Additionally, FWHM of Sk-I decreases more significantly compared to Sk-II, approaching close level at $B_\perp$ = 0.4 T [Fig. 4(i)].

It is worth noting that, the stronger $\Delta f_{sk}$ of Sk-I indicates a deeper magnetized structure extending along z-direction, typically the entire sample [10]. In contrast, the weaker $\Delta f_{sk}$ of Sk-II suggests a surface-localized magnetic structure. Moreover, the shrinking core area of Sk-I under increasing $B_\perp$ reflects a bubble-like structure in the xy-plane [37,38], whereas Sk-II maintains a nearly consistent size, indicating a pure skyrmion state. A previous theoretical study proposed a chiral bobber model of the skyrmion, which exists at the surface of the sample and aligns with the Sk-II observed in this work [39]. Subsequent experimental studies have reported the coexistence of skyrmionic bobbers and skyrmionic tubes using multiple techniques [40,41]. More recently, skyrmionic bobbers have been identified and distinguished from skyrmionic tubes based on MFM signal contrast [42,43]. Building on these prior reports and the strong experimental signatures observed in this study, we propose that Sk-I corresponds to a skyrmionic tube structure, while Sk-II corresponds to a skyrmionic bobber. More detailed discussion of three-dimensional structures of skyrmions under in-plane magnetic field is included in Supplementary Material.

## V. DISCUSSION AND SIMULATION

Our findings highlight the critical role of a strong in-plane magnetic field $B_\parallel$* in generating small branching structures and following high-density Sk-II skyrmions, whereas a lack of strong $B_\parallel$* results in Sk-I skyrmion generation [Fig. 4(j)]. The understanding of $B_\parallel$* dependent phenomena begins with the Bloch-type domain wall in $Fe_3GaTe_2$ bulk, confirmed by the mutually perpendicular relationship between the stripes domains and $B_\parallel$' (step ⑤). The Bloch-type chirality is protected and restricted by bulk DMI [44], defined as $H_{DMI}$ = -$D_{ij}$·($m_i \times m_j$), which allows the $B_\parallel$ to modulate the propagation period of stripe domains and consequently squeezing out small branching structures and sandwich skyrmion chains from the stripe domains.

The applied $B_\parallel$ introduces a parallel component $m_\parallel$ to the magnetic moment $m$, resulting in the transition of Fe$_3$GaTe$_2$ bulk magnetic structure from a helical phase to a conical phase [Fig. 4(k)] [2]. Under a strong in-plane magnetic field $B_\parallel^*$ along the x direction, the system's original vertical magnetic moment $m$ tilts toward the field direction, reducing its vertical component $m_v$ in the yz-plane [45,46]. In steady-state conditions, the rotation of $m_v$ perpendicular to $B_\parallel$ provides a non-zero cross-product term ($m_{vi} \times m_{vj}$) for DMI energy. As the magnitude of $m_v$ decreases, maintaining consistent DMI energy requires $m_v$ to rotate with a higher angular velocity in space, resulting in a shorter propagation period of stripe domains. When $B_\parallel^*$ becomes sufficiently strong to double the angular velocity of $m_v$, new magnetic domains emerge between two existing stripes. The formation mechanisms are strongly supported by the experimental measurements of $B_\parallel$ dependent $m_\perp$ variations [Fig. 3(h)].

Subsequently, in step ⑦, as $B_\parallel$ is reduced to zero, the in-plane component of $m$ vanishes, driving the system from the conical phase back to the helical phase, while the newly formed fractional magnetic domains persist due to topological protection. Finally, with the increase of $B_\perp$ in step ⑧, magnetic domains with inverse $m_\perp$ are eroded, thereby transforming the fractional magnetic domains into skyrmion chains.

We further confirmed the formation mechanism of STA by performing micromagnetic simulations based on the Landau–Lifshitz–Gilbert (LLG) equation, considering multiple factors including the Zeeman effect, magnetocrystalline anisotropy, Heisenberg exchange, DMI, demagnetization, and finite temperature.

The initial state of the simulation is set as highly oriented stripe domains with an out-of-plane magnetic field $B_\perp$ [Fig. 4(l)], corresponding to stipe phase observed experimentally in steps ④ and ⑤. When applying an in-plane magnetic field $B_\parallel$, additional signals gradually emerge, showing less prominent stripe patterns between the initial stripes due to the shortened propagation period, corresponding to experimental patterns in step ⑥. Simulating the magnetic operation in steps ⑦ and ⑧ by withdrawing the in-plane magnetic field and restoring the out-of-plane magnetic field, successfully restore both the propagation period and the initial stripe pattern. The less prominent stripe patterns gradually transform into fragmental structures and finally into skyrmion chains between the initial stripes, reproducing the highly oriented STA structure in the simulation. Furthermore, micromagnetic simulations confirmed the Bloch-type nature of skyrmions in STA, consistent with the Bloch-type domain wall behavior observed in stripe domains [Figs. S7d-e]. The full evolution process of magnetic domains in simulation is shown in details in a movie (see Supplementary Material movie S2).

Careful analysis of experimental results and simulations reveals several key magnetic properties of $Fe_3GaTe_2$ that play a role in the formation of STA, including DMI, anisotropy, and demagnetizing energy. External factors, such as temperature, also exert an influence. We propose that by deliberately controlling these factors, STA could be extended to a broader range of magnetic materials and spintronic devices (see Supplementary Material "Generalization of STA" and "Temperature dependence of STA formation").

In conclusion, we reported the generation of a large-area, highly oriented STA in ferromagnet $Fe_3GaTe_2$. The orientation and ordering of STA, as well as the density and types of skyrmions are all efficiently manipulated by our strategy with vector magnetic field operations. We examined the behavioral differences and formation processes of two distinct skyrmion types, finding that the unique formation mechanism of type II skyrmions in STA is closely tied to the interplay between DMI and the in-plane magnetic field, as supported by simulation results. This approach for generating ordered magnetic structures with high skyrmion density holds promising potential for application in skyrmion-based spintronic devices, where straight stripe domains may serve as racetracks to confine current-driven skyrmion motions in further attempts.

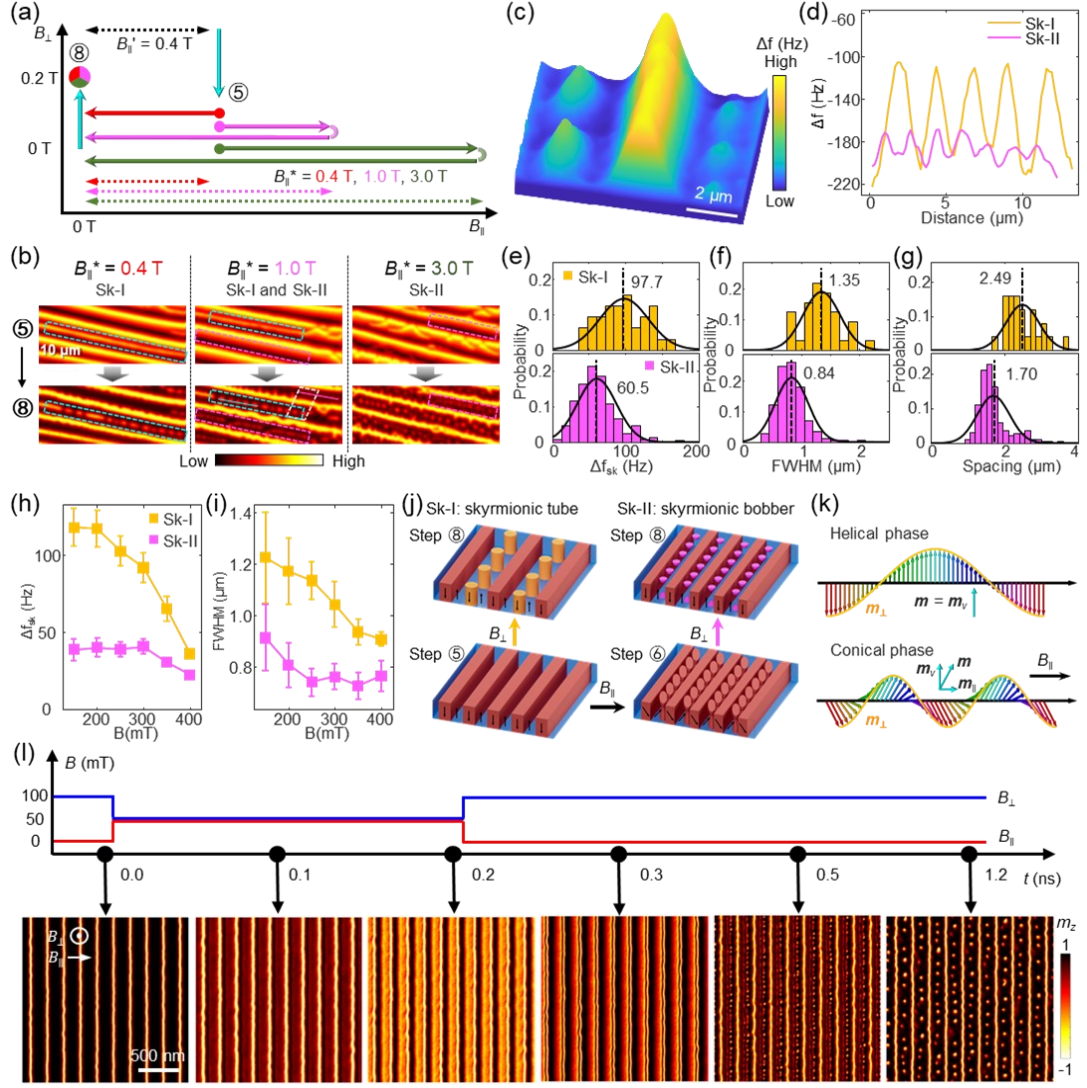

FIG. 4. **Coexistence of two types of skyrmions and mechanism of generation of STA**. (a) Schematics of generation process of STA under different $B_\parallel^*$. (b) Magnetic domains in step ⑤ ($B_\perp = 0$ T, $B_\parallel = 0.4$ T) and step ⑧ ($B_\perp = 0.2$ T, $B_\parallel = 0$ T) induced by different $B_\parallel^*$ in (a), revealing two distinct skyrmion formation mechanism. (c) Three-dimensional $\Delta f$ signal map in the white dashed box in (b). (d) Linecuts revealing $\Delta f_{sk}$ contrast between type I skyrmions (Sk-I, orange line) and type II skyrmions (Sk-II, purple line), marked by cyan and purple dashed lines in (b), respectively. (e)-(g) $\Delta f_{sk}$, FWHM and spacing distribution of two types of skyrmions in step ⑧ ($B_\perp = 0.2$ T, $B_\parallel = 0$ T). (h),(i) Averaged $\Delta f_{sk}$ and FWHM of two types of skyrmions under increasing $B_\perp$ ($B_\parallel = 0$ T). (j) Schematics of magnetic domain evolution for two types of skyrmion generation mechanism. (k) Side view (xz-plane) diagram of magnetic moment distribution in helical phase and conical phase under in-plane magnetic field $B_\parallel$ (x-

direction). $m_v$ and $m_∥$ represents the component of magnetic moment $m$ vertical and parallel to the in-plane field direction. $m_⊥$ represents the out-of-plane component of $m$. (l) Micromagnetic simulation of generation process of STA, under combined appliance of out-of-plane and in-plane magnetic field.

## Acknowledgements

The work is supported by grants from the National Natural Science Foundation of China (62488201 and 12374199), the National Key Research and Development Projects of China (2022YFA1204100), the Beijing Nova Program (Nos. 20240484651), and the Innovation Program of Quantum Science and Technology (2021ZD0302700).

## References

[1] A. Fert, V. Cros, and J. Sampaio, Skyrmions on the track, Nat. Nanotechnol. **8**, 152 (2013).

[2] S. Mühlbauer, B. Binz, F. Jonietz, C. Pfleiderer, A. Rosch, A. Neubauer, R. Georgii, and P. Böni, Skyrmion Lattice in a Chiral Magnet, Science **323**, 915 (2009).

[3] X. Z. Yu, Y. Onose, N. Kanazawa, J. H. Park, J. H. Han, Y. Matsui, N. Nagaosa, and Y. Tokura, Real-space observation of a two-dimensional skyrmion crystal, Nature **465**, 901 (2010).

[4] S. Seki, X. Z. Yu, S. Ishiwata, and Y. Tokura, Observation of Skyrmions in a Multiferroic Material, Science **336**, 198 (2012).

[5] T. Kurumaji, T. Nakajima, M. Hirschberger, A. Kikkawa, Y. Yamasaki, H. Sagayama, H. Nakao, Y. Taguchi, T. Arima, and Y. Tokura, Skyrmion lattice with a giant topological Hall effect in a frustrated triangular-lattice magnet, Science **365**, 914 (2019).

[6] L. Powalla et al., Seeding and Emergence of Composite Skyrmions in a van der Waals Magnet, Adv. Mater. **35**, 2208930 (2023).

[7] W. Jiang et al., Blowing magnetic skyrmion bubbles, Science **349**, 283 (2015).

[8] V. T. Pham et al., Fast current-induced skyrmion motion in synthetic antiferromagnets, Science **384**, 307 (2024).

[9] A. Soumyanarayanan et al., Tunable room-temperature magnetic skyrmions in Ir/Fe/Co/Pt multilayers, Nat. Mater. **16**, 898 (2017).

[10] A.-O. Mandru, O. Yıldırım, R. Tomasello, P. Heistracher, M. Penedo, A. Giordano,

D. Suess, G. Finocchio, and H. J. Hug, Coexistence of distinct skyrmion phases observed in hybrid ferromagnetic/ferrimagnetic multilayers, Nat. Commun. **11**, 6365 (2020).

[11] H. Du, J. P. DeGrave, F. Xue, D. Liang, W. Ning, J. Yang, M. Tian, Y. Zhang, and S. Jin, Highly Stable Skyrmion State in Helimagnetic MnSi Nanowires, Nano Lett. **14**, 2026 (2014).

[12] N. Mathur, F. S. Yasin, M. J. Stolt, T. Nagai, K. Kimoto, H. Du, M. Tian, Y. Tokura, X. Yu, and S. Jin, In-Plane Magnetic Field-Driven Creation and Annihilation of Magnetic Skyrmion Strings in Nanostructures, Adv. Funct. Mater. **31**, 2008521 (2021).

[13] C. Gong et al., Discovery of intrinsic ferromagnetism in two-dimensional van der Waals crystals, Nature **546**, 265 (2017).

[14] B. Huang et al., Layer-dependent ferromagnetism in a van der Waals crystal down to the monolayer limit, Nature **546**, 270 (2017).

[15] G. Zhang, F. Guo, H. Wu, X. Wen, L. Yang, W. Jin, W. Zhang, and H. Chang, Above-room-temperature strong intrinsic ferromagnetism in 2D van der Waals $Fe_3GaTe_2$ with large perpendicular magnetic anisotropy, Nat. Commun. **13**, 5067 (2022).

[16] W. Li et al., Room-Temperature van der Waals Ferromagnet Switching by Spin-Orbit Torques, Adv. Mater. **35**, 2303688 (2023).

[17] Y. Ji et al., Direct Observation of Room-Temperature Magnetic Skyrmion Motion Driven by Ultra-Low Current Density in Van Der Waals Ferromagnets, Adv. Mater. **36**, 2312013 (2024).

[18] Z. Li et al., Room-temperature sub-100 nm Néel-type skyrmions in non-stoichiometric van der Waals ferromagnet $Fe_{3-x}GaTe_2$ with ultrafast laser writability, Nat. Commun. **15**, 1017 (2024).

[19] C. Zhang et al., Above-room-temperature chiral skyrmion lattice and Dzyaloshinskii–Moriya interaction in a van der Waals ferromagnet $Fe_{3-x}GaTe_2$, Nat. Commun. **15**, 4472 (2024).

[20] H. Zhang et al., Spin disorder control of topological spin texture, Nat. Commun. **15**, 3828 (2024).

[21] X. Lv et al., Distinct skyrmion phases at room temperature in two-dimensional ferromagnet $Fe_3GaTe_2$, Nat. Commun. **15**, 3278 (2024).

[22] G. Hu et al., Room-Temperature Antisymmetric Magnetoresistance in van der Waals Ferromagnet $Fe_3GaTe_2$ Nanosheets, Adv. Mater. **36**, 2403154 (2024).

[23] H. Shi et al., Dynamic Behavior of Above-Room-Temperature Robust Skyrmions in 2D Van der Waals Magnet, Nano Lett. acs.nanolett.4c02835 (2024).

[24] S. Woo et al., Observation of room-temperature magnetic skyrmions and their current-driven dynamics in ultrathin metallic ferromagnets, Nat. Mater. **15**, 501 (2016).

[25] H. Zhang et al., Room-temperature skyrmion lattice in a layered magnet

(Fe$_{0.5}$Co$_{0.5}$)$_5$GeTe$_2$, Sci. Adv. **8**, eabm7103 (2022).

[26] S.-G. Je et al., Creation of Magnetic Skyrmion Bubble Lattices by Ultrafast Laser in Ultrathin Films, Nano Lett. **18**, 7362 (2018).

[27] Y. Guang et al., Creating zero-field skyrmions in exchange-biased multilayers through X-ray illumination, Nat. Commun. **11**, 949 (2020).

[28] Y. Guang et al., Electron Beam Lithography of Magnetic Skyrmions, Adv. Mater. **32**, 2003003 (2020).

[29] Z. Li et al., Electron-Assisted Generation and Straight Movement of Skyrmion Bubble in Kagome TbMn$_6$Sn$_6$, Adv. Mater. **36**, 2309538 (2024).

[30] N. Romming, C. Hanneken, M. Menzel, J. E. Bickel, B. Wolter, K. Von Bergmann, A. Kubetzka, and R. Wiesendanger, Writing and Deleting Single Magnetic Skyrmions, Science **341**, 636 (2013).

[31] W. Jiang et al., Direct observation of the skyrmion Hall effect, Nat. Phys. **13**, 162 (2017).

[32] G. Yu et al., Room-Temperature Skyrmion Shift Device for Memory Application, Nano Lett. **17**, 261 (2017).

[33] C. Song, L. Zhao, J. Liu, and W. Jiang, Experimental Realization of a Skyrmion Circulator, Nano Lett. **22**, 9638 (2022).

[34] Z. He et al., Experimental observation of current-driven antiskyrmion sliding in stripe domains, Nat. Mater. **23**, 1048 (2024).

[35] Z. Li et al., Field-free topological behavior in the magnetic domain wall of ferrimagnetic GdFeCo, Nat. Commun. **12**, 5604 (2021).

[36] H. Du et al., Edge-mediated skyrmion chain and its collective dynamics in a confined geometry, Nat. Commun. **6**, 8504 (2015).

[37] X. Yu, M. Mostovoy, Y. Tokunaga, W. Zhang, K. Kimoto, Y. Matsui, Y. Kaneko, N. Nagaosa, and Y. Tokura, Magnetic stripes and skyrmions with helicity reversals, Proc. Natl. Acad. Sci. **109**, 8856 (2012).

[38] M.-G. Han, J. A. Garlow, Y. Liu, H. Zhang, J. Li, D. DiMarzio, M. W. Knight, C. Petrovic, D. Jariwala, and Y. Zhu, Topological Magnetic-Spin Textures in Two-Dimensional van der Waals Cr$_2$Ge$_2$Te$_6$, Nano Lett. **19**, 7859 (2019).

[39] F. N. Rybakov, A. B. Borisov, S. Blügel, and N. S. Kiselev, New Type of Stable Particlelike States in Chiral Magnets, Phys. Rev. Lett. **115**, 117201 (2015).

[40] F. Zheng et al., Experimental observation of chiral magnetic bobbers in B20-type FeGe, Nat. Nanotechnol. **13**, 451 (2018).

[41] K. Ran, Y. Liu, Y. Guang, D. M. Burn, G. van der Laan, T. Hesjedal, H. Du, G. Yu, and S. Zhang, Creation of a Chiral Bobber Lattice in Helimagnet-Multilayer Heterostructures, Phys Rev Lett **126**, 017204 (2021).

[42] M. Grelier, F. Godel, A. Vecchiola, S. Collin, K. Bouzehouane, A. Fert, V. Cros, and N. Reyren, Three-dimensional skyrmionic cocoons in magnetic multilayers, Nat. Commun. **13**, 6843 (2022).

[43] A. K. Gopi, A. K. Srivastava, A. K. Sharma, A. Chakraborty, S. Das, H. Deniz, A. Ernst, B. K. Hazra, H. L. Meyerheim, and S. S. P. Parkin, Thickness-Tunable Zoology of Magnetic Spin Textures Observed in $Fe_5GeTe_2$, ACS Nano **18**, 5535 (2024).

[44] A. Chakraborty et al., Magnetic Skyrmions in a Thickness Tunable 2D Ferromagnet from a Defect Driven Dzyaloshinskii–Moriya Interaction, Adv. Mater. **34**, 2108637 (2022).

[45] J. Masell, X. Yu, N. Kanazawa, Y. Tokura, and N. Nagaosa, Combing the helical phase of chiral magnets with electric currents, Phys. Rev. B **102**, 180402 (2020).

[46] V. Ukleev, Y. Yamasaki, O. Utesov, K. Shibata, N. Kanazawa, N. Jaouen, H. Nakao, Y. Tokura, and T. Arima, Metastable solitonic states in the strained itinerant helimagnet FeGe, Phys. Rev. B **102**, 014416 (2020).